\documentclass[twocolumn]{aastex62}

\usepackage{multirow}
\usepackage{natbib}
\usepackage{microtype}

\makeatletter

\def\one{111$^{\circ}$}
\def\two{119$^{\circ}$}

\def\macs1149{MACS\,J1149.5+2223}

\def\chisq1radioactive{27202.565}
\def\params1radioactive{13}
\def\aic1radioactive{27228.565}

\newcommand{\Cantabria}{IFCA, Instituto de F\'isica de Cantabria (UC-CSIC), Av. de Los Castros s/n, 39005 Santander, Spain}
\newcommand{\IFCA}{\Cantabria}

\newcommand{\UCLA}{Department of Physics and Astronomy, University of California, Los Angeles, CA 90095}

\newcommand{\RCEU}{Research Center for the Early Universe, University of Tokyo, 7-3-1 Hongo, Bunkyo-ku, Tokyo 113-0033, Japan}
\newcommand{\TokyoPhys}{Department of Physics, University of Tokyo, 7-3-1 Hongo, Bunkyo-ku, Tokyo 113-0033, Japan}
\newcommand{\IPMU}{Kavli Institute for the Physics and Mathematics of the Universe (Kavli IPMU, WPI), University of Tokyo, 5-1-5 Kashiwanoha, Kashiwa, Chiba 277-8583, Japan}

\newcommand{\DARK}{DARK, Niels Bohr Institute, University of Copenhagen, Lyngbyvej 2, DK-2100 Copenhagen, Denmark} 
\newcommand{\DAWN}{The Cosmic Dawn Center (DAWN), Niels Bohr Institute, University of Copenhagen, Lyngbyvej 2, DK-2100 Copenhagen \O, Denmark; DTU-Space, Technical University of Denmark, Elektrovej 327, DK-2800 Kongens Lyngby, Denmark}

\newcommand{\Berkeley}{Department of Astronomy, University of California, Berkeley, CA 94720-3411, USA}
\newcommand{\Miller}{Miller Senior Fellow, Miller Institute for Basic Research in Science, University of California, Berkeley, CA  94720}

\newcommand{\UMN}{School of Physics and Astronomy, University of Minnesota, 116 Church Street SE, Minneapolis, MN 55455, USA}
\newcommand{\Arizona}{Department of Astronomy, University of Arizona, Tucson, AZ 85721, USA}

\newcommand{\UCSC}{Department of Astronomy and Astrophysics, University of California, Santa Cruz, CA 95064, USA}

\newcommand{\Hillsdale}{Department of Physics, Hillsdale College, 33 E. College St., Hillsdale, MI 49242, USA}
\newcommand{\BenGurion}{Physics Department, Ben-Gurion University of the Negev, P.O. Box 653, Beer-Sheva 8410501, Israel}
\newcommand{\EcoleNormaleSuperieure}{D\'epartement de Physique, \'Ecole Normale Sup\'erieure, Paris}
\newcommand{\Basque}{Department of Theoretical Physics, University of the Basque Country UPV/EHU, 48080 Bilbao, Spain}
\newcommand{\Donostia}{Donostia International Physics Center (DIPC), 20018 Donostia, Spain}
\newcommand{\Ikerbasque}{Ikerbasque, Basque Foundation for Science, E-48011 Bilbao, Spain}

\graphicspath{{./}{figures/}}

\submitjournal{ApJ}

\shorttitle{Extremely Magnified Star}
\shortauthors{Chen et al.}

\begin{document}



\title{Searching for Highly Magnified Stars at Cosmological Distances: Discovery of a Redshift 0.94 Blue Supergiant in Archival Images of the Galaxy Cluster MACS\,J0416.1-2403}

\newcounter{affilct}
\setcounter{affilct}{0}

\makeatletter
\newcommand{\affilref}[1]{%
  \@ifundefined{c@#1}%
    {\newcounter{#1}%
     \setcounter{#1}{\theaffilct}%
     \refstepcounter{affilct}%
     \label{#1}%
     }{}%
  \ref{#1}%
 }
\makeatother

\makeatletter
\newcommand*\affilreftxt[2]{%
  \@ifundefined{c@#1txt}
    {\newcounter{#1txt}%
     \setcounter{#1txt}{1}
     \altaffiltext{\ref{#1}}{#2}
     }{
     }
  }
\makeatother


\correspondingauthor{Wenlei Chen}

\author{Wenlei Chen}
\affil{\UMN}
\author{Patrick L. Kelly}
\affil{\UMN}
\author{Jose M. Diego}
\affil{\IFCA}
\author{Masamune Oguri}
\affil{\TokyoPhys}
\affil{\RCEU}
\affil{\IPMU}
\author{Liliya L. R. Williams}
\affil{\UMN}
\author{Adi Zitrin}
\affil{\BenGurion}
\author{Tommaso L. Treu}
\affil{\UCLA}
\author{Nathan Smith}
\affil{\Arizona}
\author{Thomas J. Broadhurst}
\affil{\Basque}
\affil{\Donostia}
\affil{\Ikerbasque}
\author{Nick Kaiser}
\affil{\EcoleNormaleSuperieure}
\author{Ryan J. Foley}
\affil{\UCSC}
\author{Alexei V. Filippenko}
\affil{\Berkeley}
\affil{\Miller}
\author{Laura Salo}
\affil{\Hillsdale}
\author{Jens Hjorth}
\affil{\DARK}
\author{Jonatan Selsing}
\affil{\DAWN}

\email{chen6339@umn.edu}




\begin{abstract} 
Individual highly magnified stars have been recently discovered at lookback times of more than half the age of the Universe, in lensed galaxies that straddle the critical curves of massive galaxy clusters. Having confirmed their detectability, it is now important to carry out systematic searches for them in order to establish their frequency, and in turn learn about the statistical properties of high-redshift stars and of the granularity of matter in the foreground deflector. Here we report the discovery of a highly magnified star at redshift $z = 0.94$ in a strongly lensed arc behind a Hubble Frontier Field galaxy cluster, MACS\,J0416.1-2403, discovered as part of a systematic archival search. The bright transient (dubbed ``Warhol'') was discovered in {\it Hubble Space Telescope} data taken on 2014 September 15 and 16. This single image faded over a period of two weeks, and observations taken on 2014 September 1 show that the duration of the microlensing event was at most four weeks in total. The light curve may also exhibit slow changes over a period of years consistent with the level of microlensing expected from stars responsible for the intracluster light (ICL) of the cluster. Optical and infrared observations taken near peak brightness can be fit by a stellar spectrum with moderate host-galaxy extinction. A blue supergiant matches the measured spectral energy distribution near peak, implying a temporary magnification of at least several thousand. While the spectrum of an O-type star would also fit the transient's spectral energy distribution, extremely luminous O-type stars are much less common than blue supergiants. The short timescale of the event and the estimated effective temperature indicate that the lensed source is an extremely magnified star.
\end{abstract}

\keywords{gravitational lensing: strong --- galaxies --- clusters: general, individual: MACS\,J0416.1-2403}


\section{Introduction} \label{sec:intro}
In 2016 May, imaging of a Hubble Frontier Field (HFF) galaxy-cluster field, MACS\,J1149.5+2223 (MACS1149; redshift $z=0.54$), revealed a several-week-long transient ($F125W (J) \approx 25.7$\,mag\,AB; $i \approx 26.4$ mag AB at peak) in a highly magnified galaxy at $z=1.49$ \citep{kellydiegorodney18}.
A highly magnified image of the lensed star has always been detected in deep {\it Hubble Space Telescope (HST)} observations,
and the spectral energy distribution (SED) of the star measured in HFF imaging in 2014 matches that of the bright transient detected in 2016 May, consistent with temporarily increased magnification.
The SED also exhibits a strong Balmer jump present in some luminous stars yet absent from stellar outbursts.  Finally, simulations of microlensing of a background star by stars or remnants in the foreground cluster can produce light curves similar to that observed \citep{diegokaiserbroadhurst18,kellydiegorodney18,venumadhavdaimiraldaescude17}. 
The discovery of the star realized a theoretical prediction that individual stars at cosmological distances could become sufficiently magnified to be detected \citep{miraldaescude91}.

\begin{figure*}
\centering
\includegraphics[angle=0,width=6.5in]{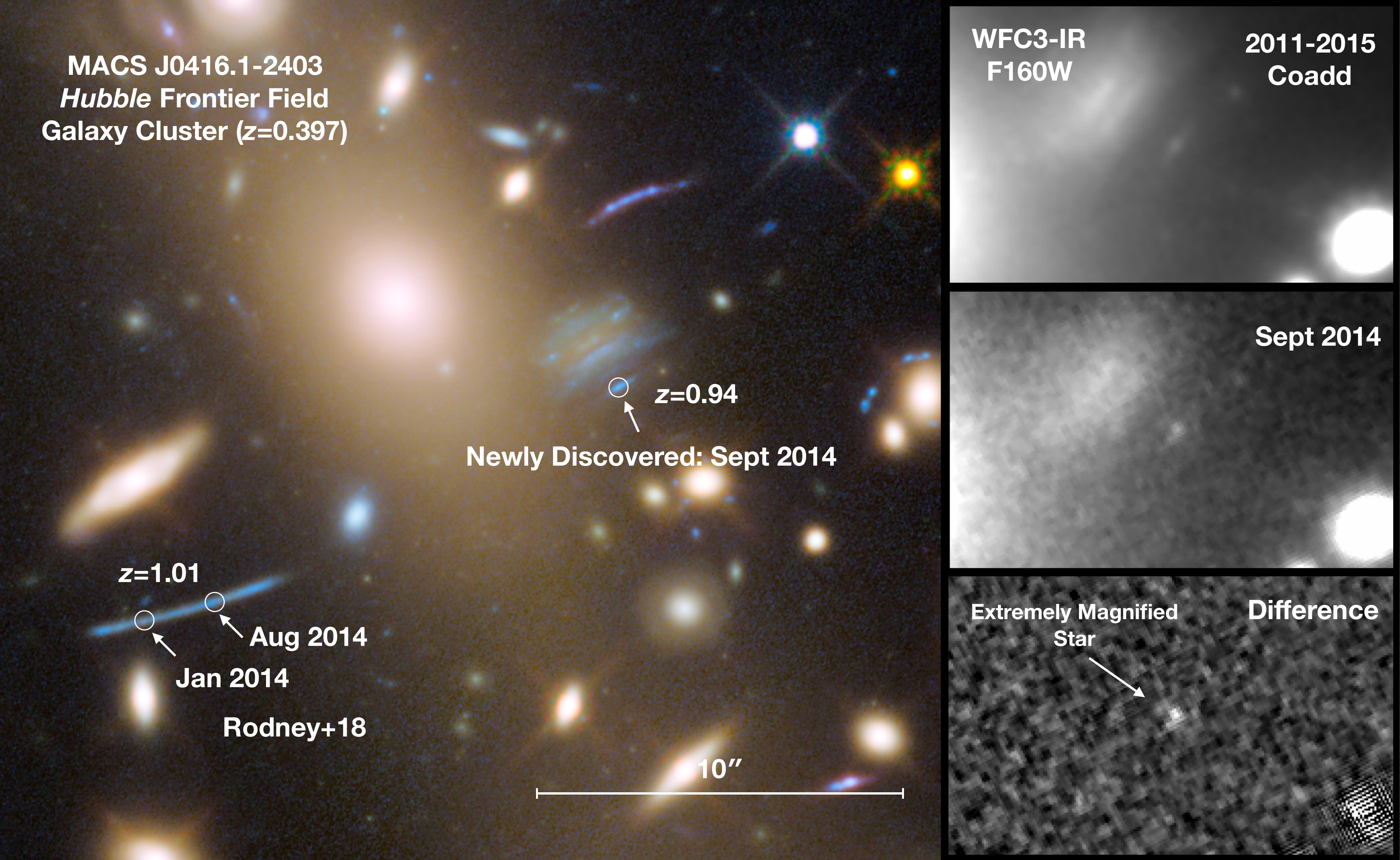}
\caption{Left panel shows the location of the newly discovered {\it extremely magnified} star in an 
arc at $z=0.94$ found in archival {\it HST} imaging of the MACS0416 galaxy cluster,
and the positions of the two stellar microlensing events previously identified by \citet{rodneybalestrabradac18} in 
a {\it different} strongly lensed galaxy at $z = 1.01$. The timescales of all three events were several 
weeks. Right panel shows an example deep template WFC3-IR F160W image of the field (top), image of the newly identified event near peak in 2014 September (middle), and the difference image (bottom). 
\label{fig:mosaic}}
\end{figure*}

\begin{figure*}
\centering
\includegraphics[angle=0,width=6.5in]{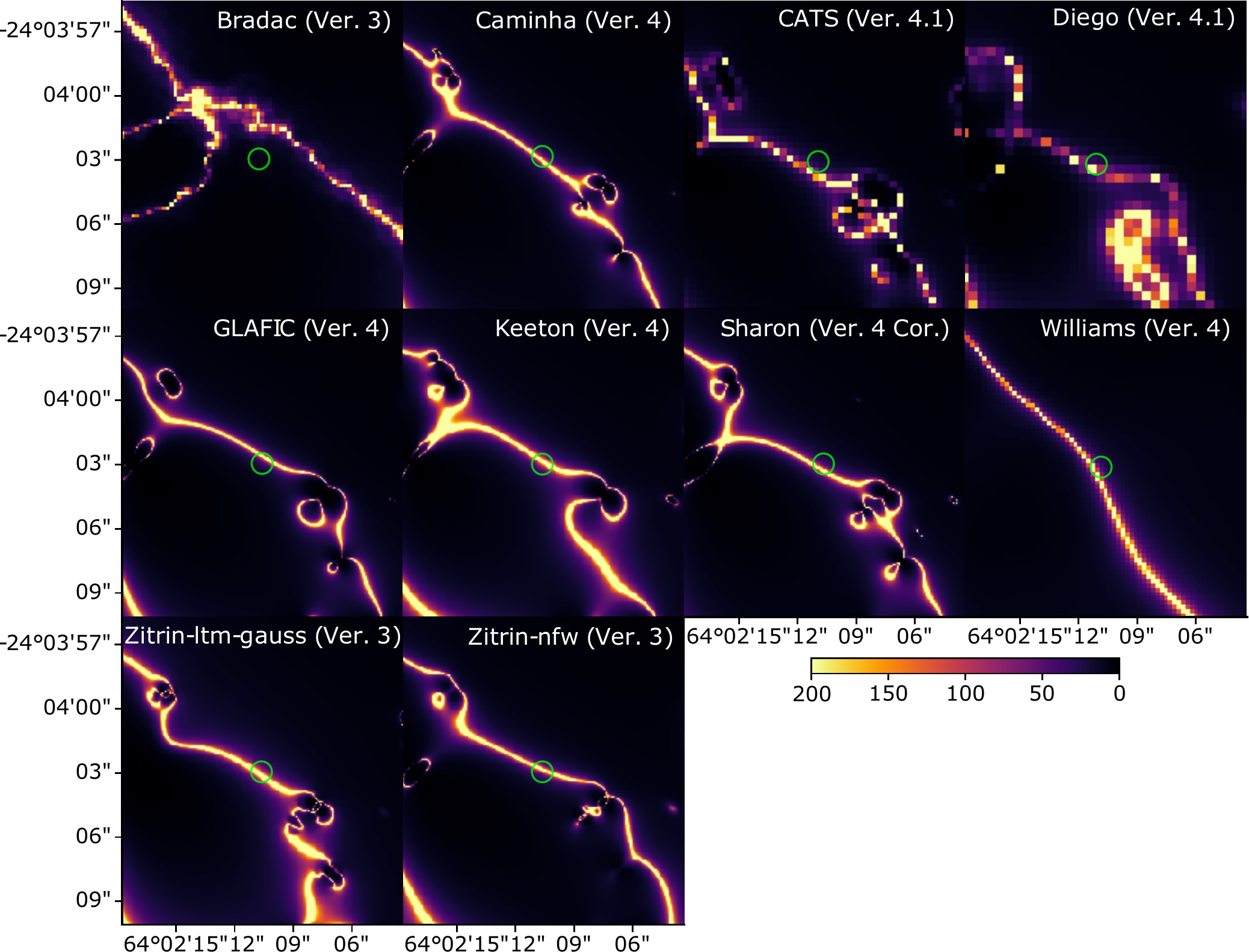}
\caption{Bright microlensing event Warhol (green circle) is close to the critical curve of the MACS0416 galaxy cluster. Panels show the magnification maps for published lens models (see Table~\ref{tab:mag} for magnification values). The galaxy-cluster critical curve has a simple configuration close to the location of the microlensing event.
\label{fig:mag_maps}}
\end{figure*}

In 2014 January and August, the FrontierSN project (PI S. Rodney) detected a pair of transients dubbed the ``Spock'' events at two separate locations
in a highly magnified galaxy at $z=1.01$ behind the MACS\,J0416.1-2403 (MACS0416; \citealt{ebelingedgehenry01}) galaxy cluster ($z=0.397$) using {\it HST}.
These events, whose locations are shown in Fig.~\ref{fig:mosaic}, were identified during two month-long campaigns to image MACS0416
as part of HFF project (PI J. Lotz). 
While the events each lasted only several weeks,  their interpretation was not immediately apparent. 
The detection of the lensed star in MACS\,J1149 magnified by $>2000$ at peak brightness prompted the interpretation of the two MACS0416 events as likely microlensing events \citep{rodneybalestrabradac18}. 


As shown in Fig.~\ref{fig:mosaic}, we have now identified a {\it third} highly magnified
star in the MACS0416 field in a {\it different} lensed galaxy at $z=0.94$ in archival
{\it HST} imaging taken in 2014 September.
We have named this transient ``Warhol" given its ``fifteen minutes of fame.'' 
Fig.~\ref{fig:mag_maps} shows that the transient is within a small fraction of an
arcsecond from the location of the MACS0416 cluster's critical curve according to published models.  
At these small separations from the critical curve, microlensing of bright stars in a background arc by objects in the foreground cluster including  stars or remnants is not only possible, but in fact inevitable.

In Section~\ref{sec:data}, we describe the imaging data in this paper. Section~\ref{sec:methods} provides the details of the methods we use to analyze the {\it HST} imaging.
In Section~\ref{sec:results}, the results of our analysis are presented, and our conclusions are given in Section~\ref{sec:summary}. All magnitudes are in the AB system \citep{okegunn83}, and we use a standard set of cosmological parameters ($\Omega_m=0.3$, $\Omega_\Lambda=0.7$,
$H_0$ = 70 km s$^{-1}$ Mpc$^{-1}$).

\section{Data}
\label{sec:data}
Imaging of the MACS0416 galaxy-cluster field with the ACS and WFC3 cameras has been acquired as part of the Cluster Lensing and Supernova survey with Hubble (CLASH; GO-12459; \citealt{postmancoebenitez12}), the
Grism Lens-Amplified Survey from Space (GLASS; PI T. Treu; GO-13459; \citealt{schmidttreubrammer14}; \citealt{treuschmidtbrammer15}), the HFF (GO-13496; \citealt{lotzkoekemoercoe17}), the FrontierSN follow-up program (PI S. Rodney; GO-13386), and the Final UV Frontier project (PI Siana; GO-14209).
Earlier imaging of the MACS0416 field, not analyzed in this paper, was acquired with the WFPC2 (PI H. Ebeling; GO-11103).
The microlensing peak we report here occurred in the target-of-opportunity imaging follow-up of the Spock events \citep{rodneybalestrabradac18} acquired by the FrontierSN program.

\begin{figure*}
\centering
\includegraphics[angle=0,width=6.5in]{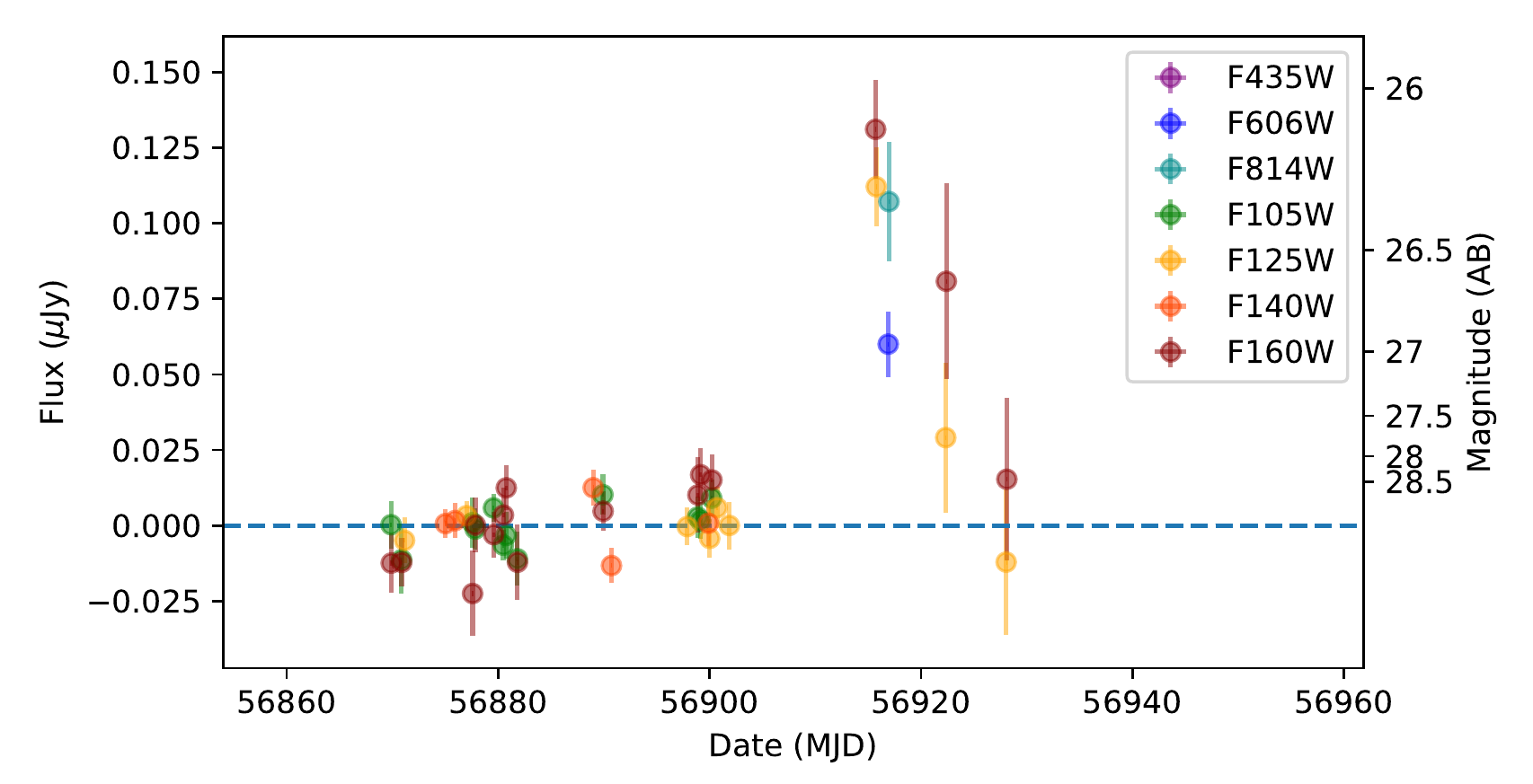}
\includegraphics[angle=0,width=6.5in]{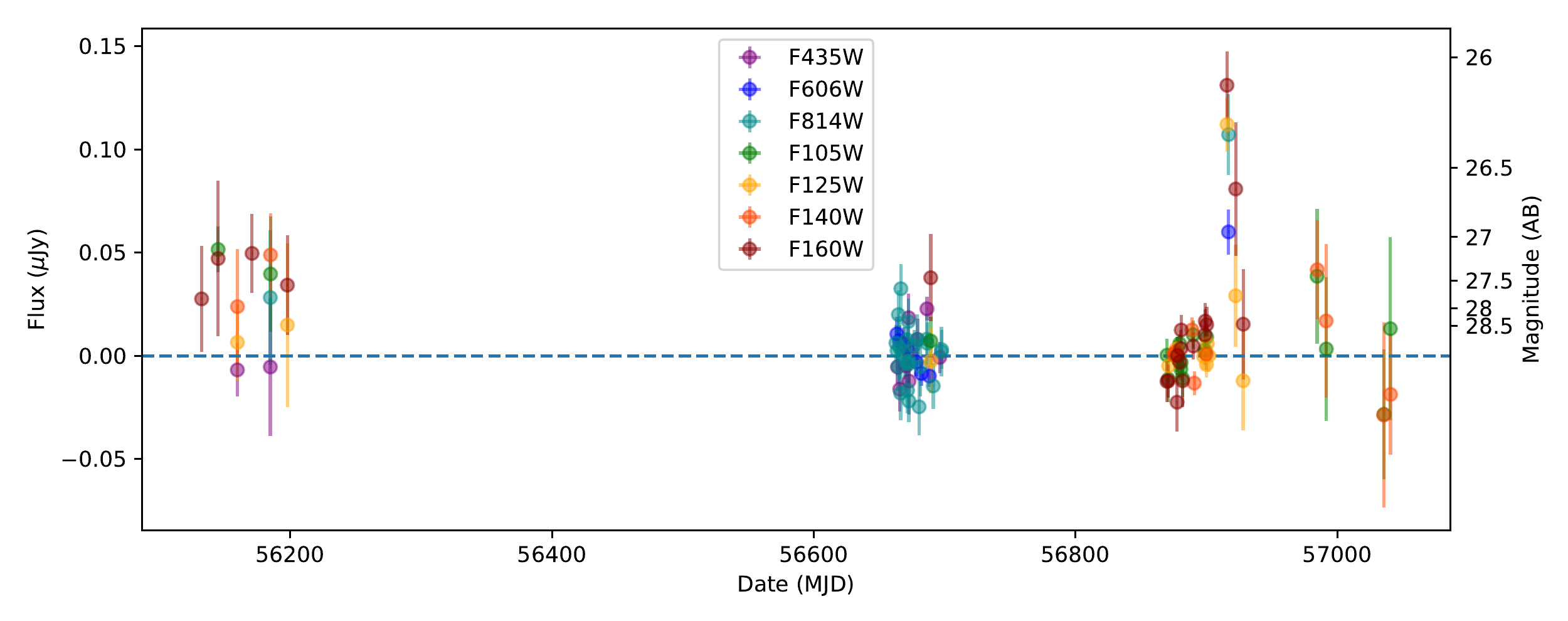}
\caption{Photometry of the newly identified microlensing event identified in archival images of the MACS0416 HFF galaxy-cluster field. The upper panel shows the multiband optical and near-infrared light curve close to peak brightness in 2014 September, and shows that its timescale is on the order of several weeks, similar to those of the microlensing events reported by \citet{kellydiegorodney18} in MACS1149 and \citet{rodneybalestrabradac18} in MACS0416.
A several-week duration is also consistent with the expected transverse velocities of galaxy clusters \citep{kellydiegorodney18,diegokaiserbroadhurst18,venumadhavdaimiraldaescude17,oguridiegokaiser18}. The lower panel plots {\it all} existing {\it HST} observations of the MACS0416 galaxy-cluster field. 
\label{fig:lightcurves}}
\end{figure*}

\begin{figure*}
\centering
\includegraphics[angle=0,width=4.7in]{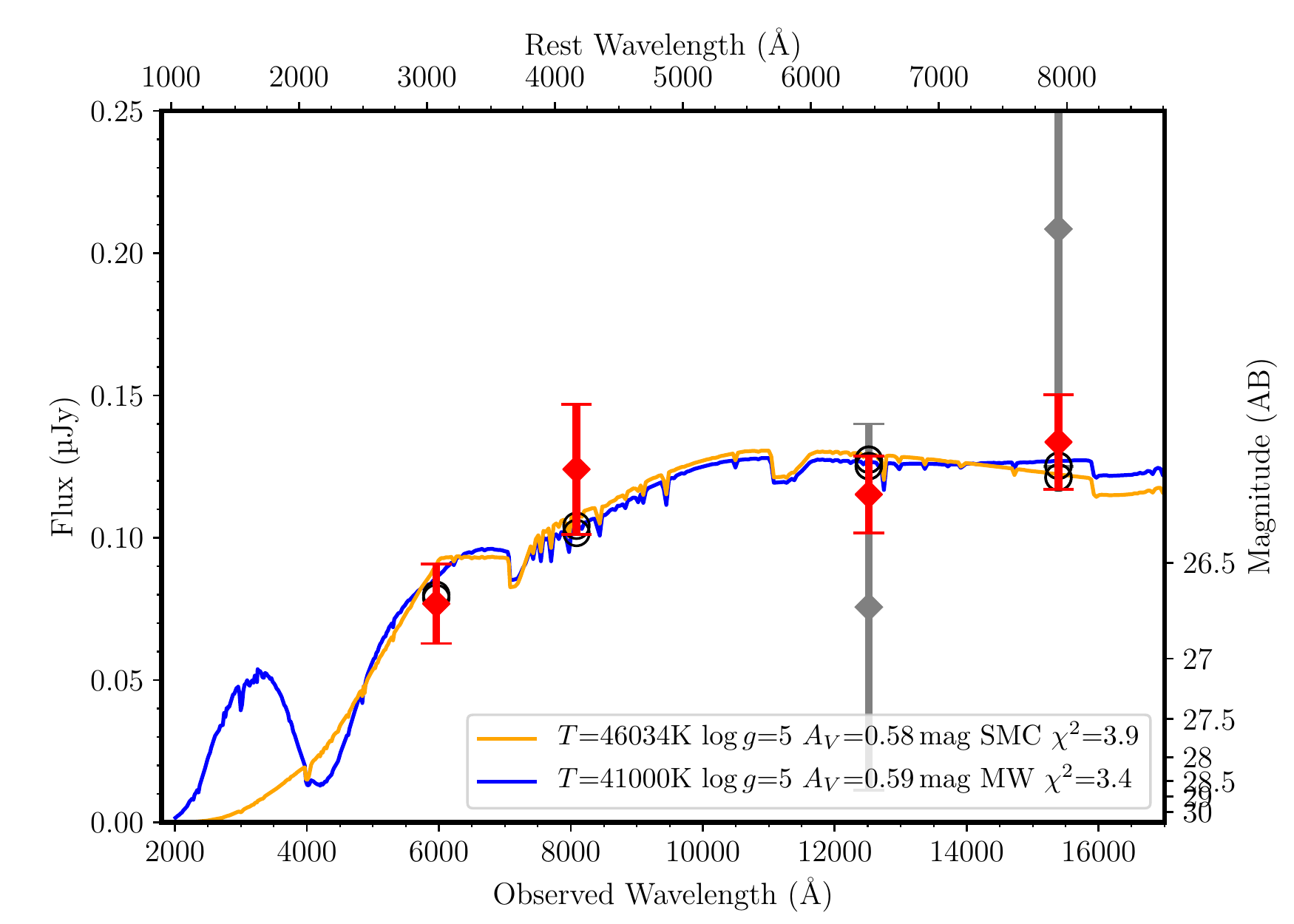}
\caption{Stellar atmosphere model with temperature $\sim40,000$\,K \citep{castellikurucz04} and host-galaxy extinction that provide the best fit to the measured SED during the 2014 September microlensing event. In this fit, we allow the temperature of the star to vary as a free parameter. The red points mark photometry measured from images taken on 2014 September 15 and 16.  The gray points mark fluxes measured from imaging acquired on 2014 September 22; given the light curve's evolution we include an additional parameter in the fit: the relative flux normalization of the event between September 15--16 and September 22.
\label{fig:sedmodelflex}}
\end{figure*}

\begin{figure*}
\centering
\includegraphics[angle=0,width=4.7in]{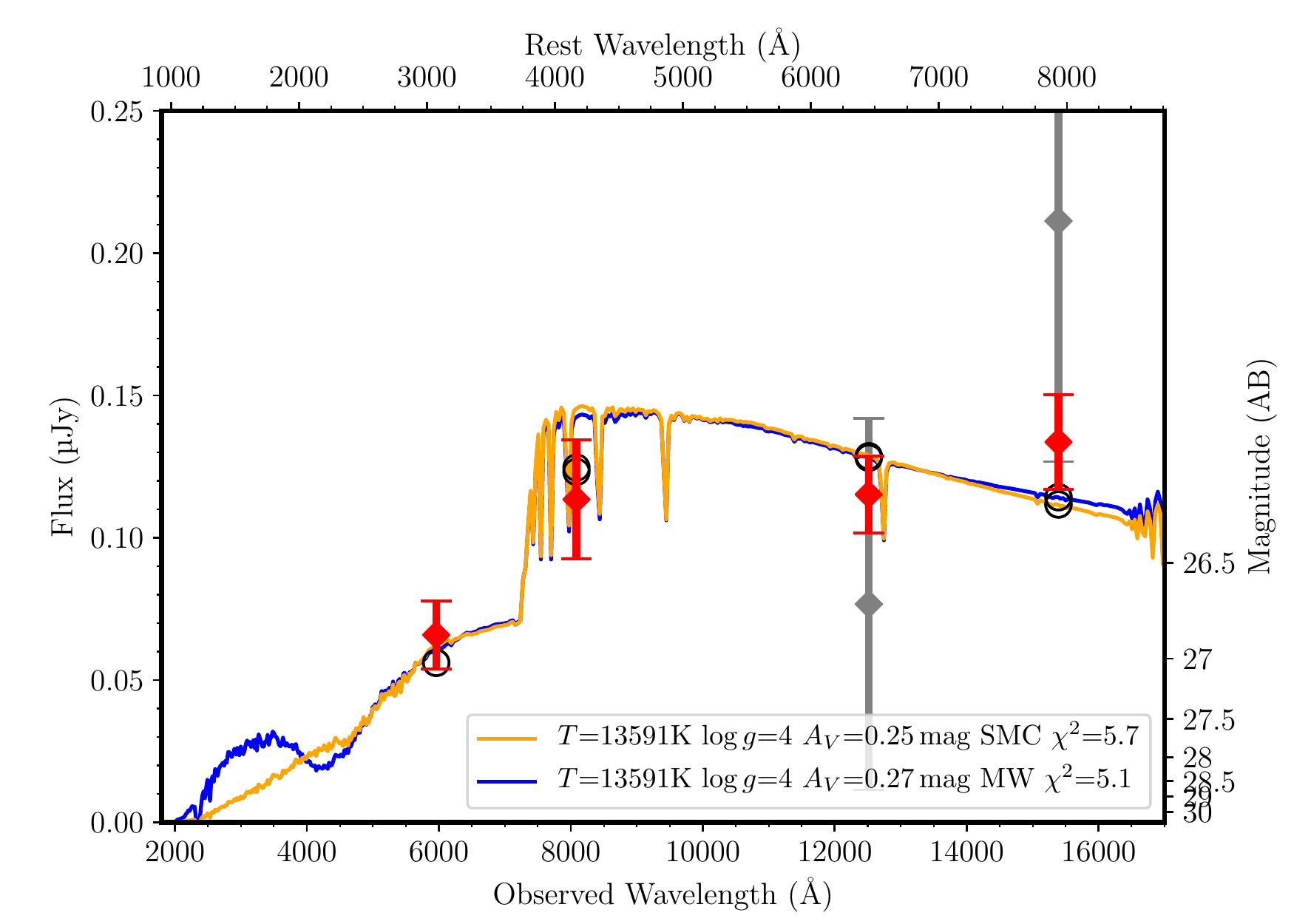}
\caption{Same as Fig.~\ref{fig:sedmodelflex}, except that here the stellar atmosphere (13,591\,K and $\log g = 4$) is one of the best-fitting models to the SED of the extremely magnified blue supergiant star Icarus in the MACS1149 field \citep{kellydiegorodney18}. Given their lower initial masses and longer lifetimes, blue supergiants luminous in the rest-frame optical are more numerous than the most massive and luminous O-type stars.
\label{fig:sedmodelfixed}}
\end{figure*}

\section{Methods}
\label{sec:methods}

\subsection{Image Processing and Coaddition}
We aligned all imaging with {\tt TweakReg}, and then resampled images to a scale of 0.03\arcsec~pixel$^{-1}$ using {\tt AstroDrizzle} \citep{fruchter10}.

\subsection{{\tt PythonPhot} Photometry}
We use {\tt PythonPhot}\footnote{\url{https://github.com/djones1040/PythonPhot}}
\citep{jonesscolnicrodney15} to measure the light curves from difference imaging.
The {\tt PythonPhot} package includes an implementation of point-spread function (PSF) fitting photometry based on the {\tt DAOPHOT} algorithm
\citep{stetson87}.

\section{Results}
\label{sec:results}

\subsection{Position and Underlying Arc}
The transient's J2000 coordinates are $\alpha = 4^{\rm h}16^{\rm m}08.7084^{\rm s}$, $\delta = -24^{\circ}04'02.945''$ in the World Coordinate System (WCS) of the official HFF coadded images. A spectrum of the underlying arc acquired by the CLASH-VLT survey yielded $z=0.93910$ \citep{balestramercuriosartoris16,caminhagrillorosati17}. The smaller redshift of the arc (compared with the previous examples of lensed stars) implies that fainter stars can be magnified above the detection threshold. \citet{patriciorichardcarton18} measure an oxygen abundance of 
12 + log(O/H) = $8.72\pm0.6$\,dex and a low extinction of $A_V = 0.15 \pm 0.20$ from nebular emission lines for the lensed system from Multi Unit Spectroscopic Explorer (MUSE) integral-field unit (IFU) spectroscopy.

\subsection{Magnification Predictions for Galaxy-Cluster Models}
We calculate magnification maps at $z=0.94$ using ten independent Frontier Fields Lens Models \citep{lotzkoekemoercoe17} for the MACS0416 galaxy cluster, as shown in Fig.~\ref{fig:mag_maps}. The predicted magnification $\mu$ due to the galaxy-cluster lens at Warhol's position is listed in Table~\ref{tab:mag}.
In general, the locations of galaxy-cluster critical curves are constrained by current models to within several tenths of an arcsecond in the best cases. 
Given Warhol's proximity to the critical curve, the uncertainty in the critical curve's location results in a large magnification uncertainty at its position.

\subsection{Light Curve and Duration of Event}
The optical and near-infrared light curve plotted in Fig.~\ref{fig:lightcurves} 
shows that the microlensing event faded over a period of at least two weeks. 
The event was at least $\sim 1.5$ times brighter (total flux) in the infrared (IR) band than its underlying arc in archival {\it HST} imaging during the HFF project, as the true peak of this microlensing event may have occurred during gaps of {\it HST} visits, as shown in Fig.~\ref{fig:lightcurves}. Photometry is measured using a 0.2\arcsec\ aperture (detailed values are listed in Table~\ref{tab:photometry}).

A microlensing peak should have a duration roughly $R/v$, where $R$ is the size of the lensed
source and $v$ is the transverse velocity of the lensing system.  Given the $\sim1000$\,km s$^{-1}$ expected relative transverse velocity between the galaxy cluster and background source, the several-week timescale of the microlensing peaks implies that 
the lensed sources can only extend for at most several tens of AU. Consequently, the lensed systems must be stellar systems (e.g., single star or binary system) instead of a star cluster. 

\subsection{A Single-Image Transient Event}
Sources near a cluster fold caustic (with no microlenses) should appear as a {\it pair} of images with equal magnification. Therefore, if the new transient were a stellar outburst, we would expect to see a pair of transients with a relative time delay of less than a day.
By contrast, a microlensing event should only appear as a single transient,
as a star or remnant in the cluster becomes temporarily aligned with one of the magnified images of the background star. As shown in Fig.~\ref{fig:warhol_ims}, only a single bright  transient along the arc was detected during the 2014 September {\it HST} visits.

 
\subsection{A Counterimage of the Lensed Star?}
\begin{figure*}
\centering
\includegraphics[angle=0,width=6.5in]{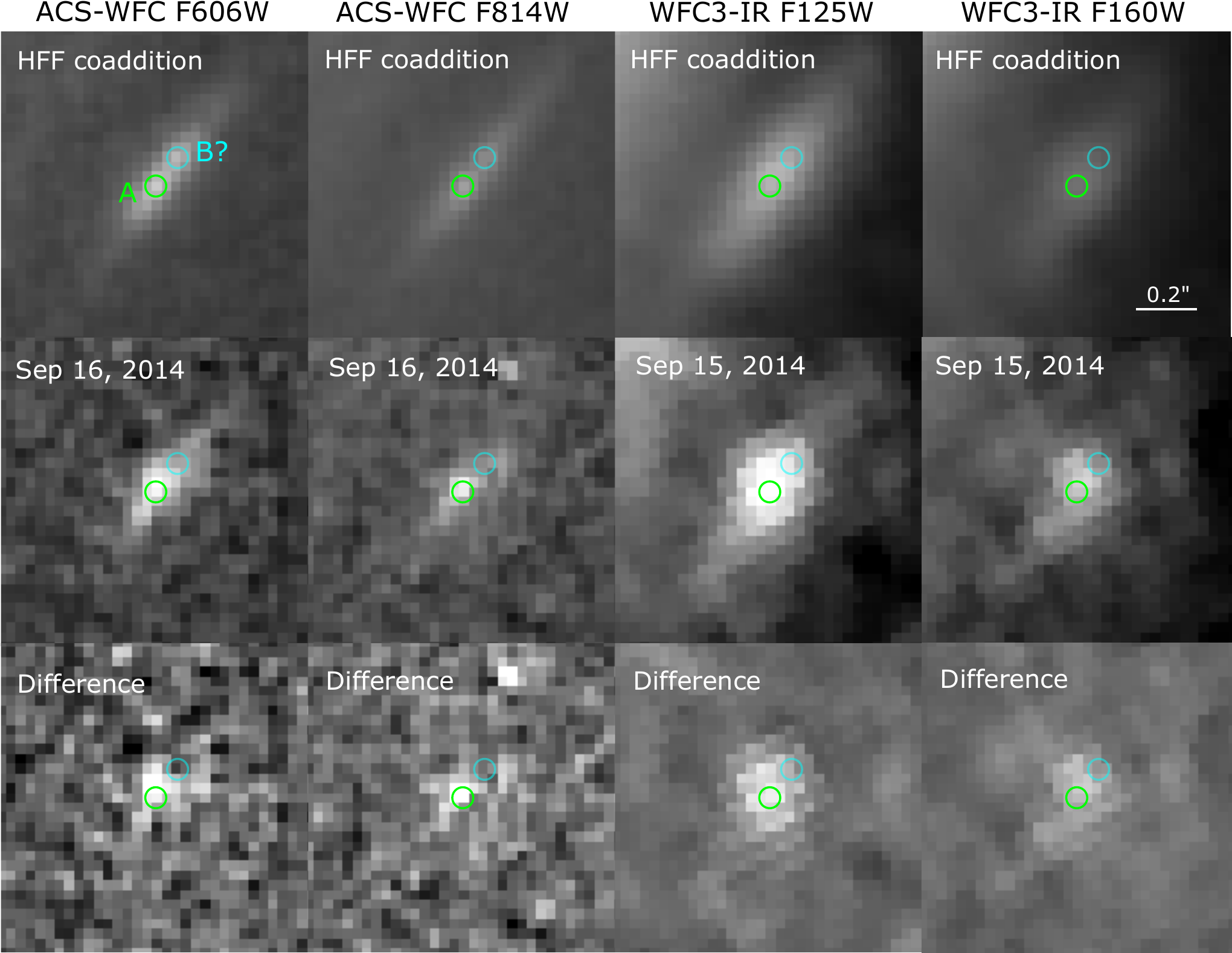}
\caption{{\it HST} imaging around Warhol's position. Upper four panels show coadditions of {\it HST} images obtained from the HFF project using ACS-WFC F606W, ACS-WFC F814W, WFC3-IR F125W, and WFC3-IR F160W (templates). Middle four panels are the {\it HST} images during the microlensing event detected around 2014 September 15. Lower four panels are the difference images. A peak (marked by the circle ``A" in the top-left panel) can be identified from the optical {\it HST} imaging in the arc. There may be another peak along the arc shown in the F606W band (as marked by the circle ``B"). The same positions of A and B are marked by green and cyan circles (respectively) in all images. Each pair of transient and template images is displayed using the same color scale.
\label{fig:warhol_ims}}
\end{figure*}

\begin{figure*}
\centering
\includegraphics[angle=0,width=6in]{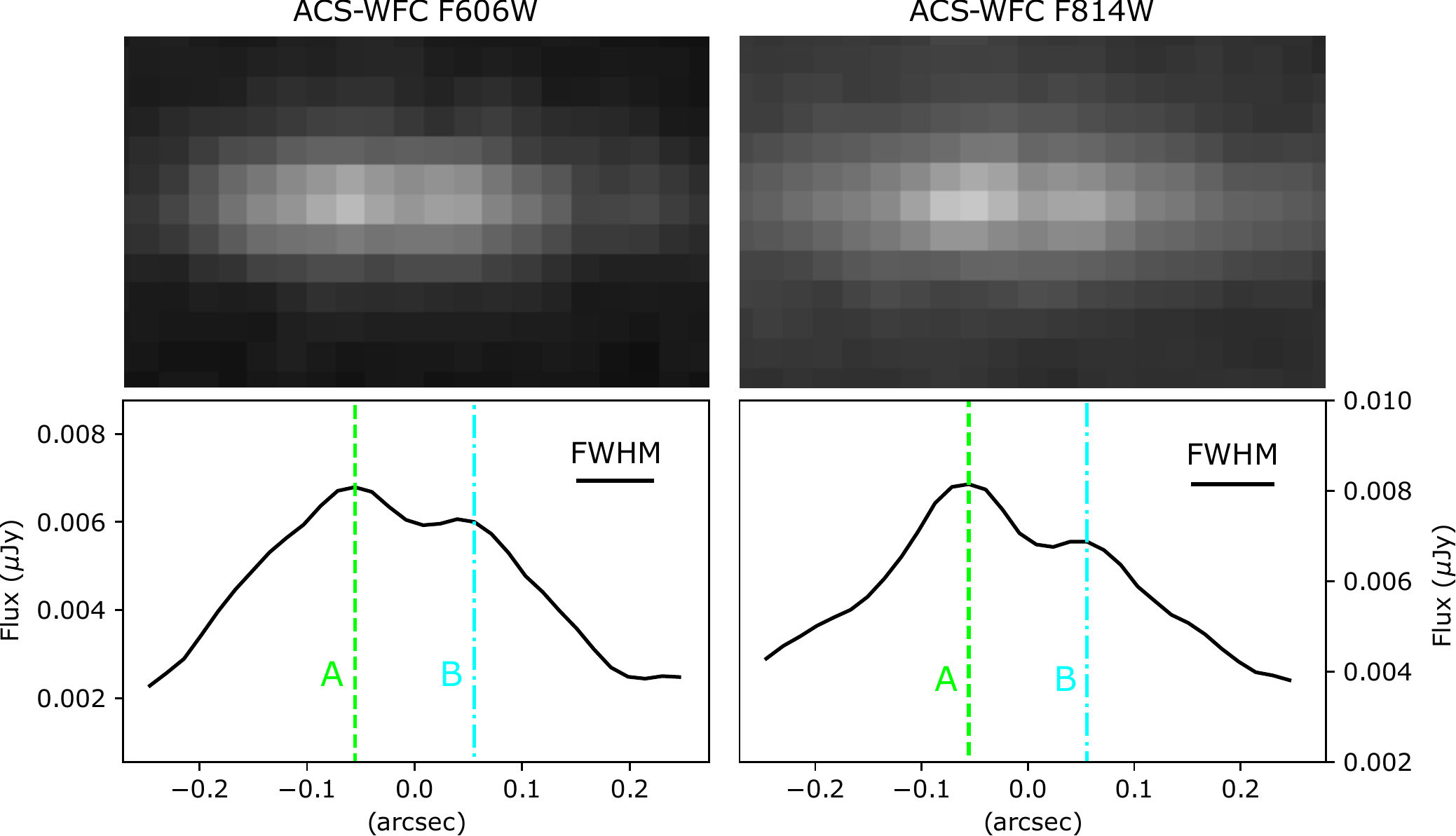}
\caption{ACS-WFC F606W and ACS-WFC F814W imaging of the underlying arc detected by the HFF project before 2014 September 15--16. Bottom two panels show flux along the arc with a 0.05\arcsec\ diameter aperture. Vertical green and cyan lines show the positions A and B (respectively) in Fig.~\ref{fig:warhol_ims}. The horizontal bars show the full width at half-maximum intensity (FWHM) of averaged PSFs in ACS-WFC F606W and ACS-WFC F814W bands.
\label{fig:arc_flux}}
\end{figure*}

Warhol's location, marked by the green circle labeled ``A" in Fig.~\ref{fig:warhol_ims}, corresponds to a peak along the underlying arc in coadditions of HFF F606W and F814W imaging acquired before the microlensing event.
To determine whether a counterimage of the underlying source may exist along the arc, we 
measured the flux inside of a 0.05\arcsec\ diameter
aperture as we moved it along the arc.
Fig.~\ref{fig:arc_flux} shows possible evidence for a second peak labeled ``B" along the underlying arc. The locations A and B are separated by $\sim0.12''$.
In the absence of microlensing, the fluxes of two counterimages should be identical. Therefore, the fact that two observed potential counterimages do not exhibit equal fluxes implies the presence of microlensing.




\begin{deluxetable*}{lcccccc}
\tablecaption{Magnification ($\mu$) required for different types of stars.\tablenotemark{a} \label{tab:stars}}
\tablecolumns{6}
\tablewidth{0pt}
\tablehead{
&
\colhead{Spec. Model} &
\colhead{Temp} &
\colhead{$M_V$} & 
\colhead{F125W} & 
\colhead{$K$} & 
\colhead{$\mu$} 
}
\startdata
BSG&B8V&11749K&-8.5&34.94&-0.52&3003\\
Extreme MS&O5V&39810K&-8&35.58&-0.39&5381\\
\hline
MS&O5V&39810K&-5.40&38.18&-0.39&59002\\
MS&O9V&35481K&-4.00&39.60&-0.37&218257\\
MS&B0V&28183K&-3.70&39.87&-0.39&280249\\
MS&B1V&22387K&-3.20&40.33&-0.43&429719\\
MS&B3V&19054K&-2.10&41.41&-0.45&1163781\\
MS&B5-7V&14125K&-2.10&41.38&-0.48&1126591\\
MS&B8V&11749K&-1.08&42.36&-0.52&2789947\\
\enddata
\tablenotetext{a}{Approximate peak magnifications are for no host-galaxy extinction and for the peak observed F125W magnitude of $\sim26.25$. ``MS'' is an abbreviation for main sequence, and ``BSG'' is an abbreviation for blue supergiant. Note that high magnifications are required for typical main-sequence stars using \citet{pickles98} templates. Consequently, we favor a post-main-sequence blue supergiant having $-9 \lesssim M_V \lesssim -7$, although an extreme and even less common O-type main-sequence star also provides a satisfactory fit to the SED (see Figs.~\ref{fig:sedmodelfixed} and \ref{fig:sedmodelflex}).} 

\end{deluxetable*}

\subsection{Spectral Energy Distribution of the Star}
After correcting for extinction expected for the Galactic foreground ($A_V = 0.112$\,mag; \citealt{schlaflyfinkbeinerSFD11}), we fit the spectral energy distribution (SED) of the microlensing peak. ACS-WFC F606W and F814W, as well as WFC-IR F125W and F160W, imaging was acquired during a first epoch on 2014 September 15--16; the optical and IR integrations were interspersed with each other in time. As shown in Fig.~\ref{fig:lightcurves}, the transient was still detected during a second imaging epoch on 2014 September 22.

We simultaneously fit a \citet{castellikurucz04} stellar atmosphere model and a host-galaxy extinction curve to the measured SED of the Warhol microlensing event.  We assume that the source did not vary significantly while the optical and IR images were acquired during the first epoch. We include as a fit parameter the change in the magnification (relative normalization of the SED) between the first and second epochs. 

Fig.~\ref{fig:sedmodelflex} shows the best fit to the measured photometry when we allow the temperature of the stellar photosphere to vary as a free parameter.  In Fig.~\ref{fig:sedmodelfixed}, we show that the best-fitting stellar model (with temperature $T \approx 13,600$\,K) to the highly magnified blue supergiant Icarus in the MACS1149 field \citep{kellydiegorodney18} also provides a reasonable fit to Warhol's SED. The low to moderate best-fitting host-galaxy dust extinction is consistent with the $A_V = 0.15 \pm 0.20$\,mag extinction inferred by \citet{patriciorichardcarton18} from an analysis of nebular emission lines from the lensed galaxy. We expect that microlensing may only potentially be chromatic when a microcaustic is close to the limb of the star, but this should occur over a very short timescale smaller than the $\sim$2 days during which observations near peak were acquired.


\subsection{Magnification}
We use the definition of a {\it K}-correction $K_{\rm xy}$ as 
\begin{equation}
m_{\rm y} = M_{\rm x} + dm + K_{\rm xy},
\end{equation}
where $m_{\rm y}$ is the observer-frame apparent magnitude in the {\rm y} band, $M_{\rm x}$ is the rest-frame absolute magnitude in the {\rm x} band, and $dm$ is the distance modulus. To calculate a {\it K}-correction, we use Eq. 2 of  \citet{kimgoobarperlmutter96},
\begin{equation}
K=2.5\,\log_{10}(1+z) + m_{F125W,{\rm syn}}^{\rm AB} - m_{V, {\rm syn}}^{\rm Vega},
\end{equation}  
where $z=0.94$, $m_{F125W,{\rm syn}}^{\rm AB}$ is the WFC3 $F125W$ synthetic magnitude of a redshifted model spectrum, and $m_{V,{\rm syn}}^{\rm Vega}$ is the synthetic Johnson {\it V}-band magnitude of the rest-frame model spectrum. 
Using the best-fitting spectral models, we calculate $K_{V,F125W} \approx -0.4$, and adopt $dm=43.96$\,mag at $z=0.94$ (with no correction for magnification).

For fold caustics, the source-plane area $A$ within which the magnification exceeds $\mu$ scales as $A(>\mu) \propto 1/\mu^2$. Consequently, low-magnification microlensing events have greater probability of occurring. 
In the case of Icarus, a persistent image of the lensed blue supergiant has always been detected in deep {\it HST} imaging, and the magnification is on average 300--600, with inferred microlensing magnification reaching up to $>2000$ during an event in 2016 \citep{kellydiegorodney18}.

The luminosities of blue supergiant stars in the Small Magellanic Cloud (SMC) reach $M_V \gtrsim -8.8$\,mag \cite{dachs70}. 
There are examples of extremely luminous, main-sequence O-type stars such as Melnick 34 in the 30 Doradus complex in the Large Magellanic Cloud (LMC) with an absolute magnitude of $M_V = -7.9$ \citep{dorancrowtherdekoter13}.

However, a very luminous O-type star or Wolf-Rayet stars showing H in their spectra (WNH) should be extremely rare, whereas cooler B-type supergiants at lower bolometric luminosity but similar $M_V$ will be much more common in the field of a star-forming galaxy. The comparatively small abundance of WNH stars arises from their initial masses (very approximately 100 $M_{\odot}$ vs. 10--20 $M_{\odot}$), but also the significantly longer lifetimes at the lower masses. Among binary stars, blue supergiants can also be blue stragglers from mass gainers and mergers.

For a blue supergiant star with identical temperature and luminosity, the $F125W \approx 26.25$ apparent magnitude observed for Warhol at $z=0.94$ would require a factor $\sim3.5$ smaller magnification than would Icarus at $z=1.49$ given its peak apparent magnitude of $F125W \approx 25.5$. We note, however, that Warhol's light curve likely does not include its peak brightness.
In Table~\ref{tab:stars}, we list the magnification required for main-sequence and post-main-sequence stars of different spectroscopic types.

\subsection{Constraints on Source Size}
The separation between A and its possible counterimage B is $\sim0.12''$.  Near the critical curve, the {\tt GLAFIC} galaxy-cluster mass model yields the following magnification for each of the images, in the case of a smooth model (i.e., with no microlensing): 
\begin{equation}
\mu_{\rm each} \approx (11\,{\rm arcsec}) / \theta_h ,
\end{equation}
where $\theta_h$ is the angular distance from the critical curve.
At an offset of $\theta_h \approx 0.06''$ from the galaxy-cluster critical curve, $\mu_{\rm each} \approx 180$ ($\mu_{t} \approx 120$, $\mu_r \approx 1.5$), which may be a plausible location for caustic crossing given the saturation argument presented by \citet{diegokaiserbroadhurst18}.

We also compute the source crossing time as
\begin{equation}
t_{\rm src} \approx 0.031 \frac{R_{\rm source} / R_{\odot}}{v / (500\,{\rm km}\, {\rm s}^{-1})}\,{\rm days},
\end{equation}
where $v$ is a transverse velocity of the cluster and $R_{\rm source}$ is a radius of a backgroud star that is magnified.
From the light curve, we have $t_{\rm src} < 10$ days (see Fig.~2 of \citealt{miraldaescude91} to see how $t_{\rm src}$ relates to the expected timescale of the light curve's evolution),
which yields a limit of $R_{\rm source} < 320 R_{\odot}$.
Assuming $\mu_{t} = 120$ and $\mu_{r} = 1.5$, the maximum magnification estimated using Eq. 48 of \citet{oguridiegokaiser18} is
\begin{equation}
\mu_{\rm max} \approx (4.5 \times 10^4) (M_{\rm lens} / M_{\odot})^{1/4}  (R_{\rm source} / R_{\odot})^{-1/2}.
\end{equation}
For $M_{\rm lens} = 0.3 M_{\odot}$ (typical mass of a star responsible for the intracluster light of the cluster), we have 
$\mu_{\rm max} \approx 33,000$ for $R_{\rm source} = 1 R_{\odot}$, 
$\mu_{\rm max} \approx 10,000$ for $R_{\rm source} = 10 R_{\odot}$,
and $\mu_{\rm max} \approx 3300$ for $R_{\rm source} = 100 R_{\odot}$,
where a larger $M_{\rm lens}$ yields a greater maximum magnification.
The comparison with Table~\ref{tab:stars} suggests that normal main-sequence stars are unlikely to be observed as microlensing events, and we need to consider either blue supergiants or extremely luminous O-type stars to explain the Warhol event.

As shown in Fig.~\ref{fig:arc_flux}, the sources A and B appear to be unresolved in HFF F606W and F814W imaging acquired before the microlensing event. An approximate estimate, assuming a transversal magnification of $\sim$100, indicates that the coincident source at positions A and B detected in HFF imaging occurred must be $\sim$3 pc at most, so it must be a single star, stellar system, or a compact stellar cluster.

\begin{deluxetable*}{lccc}
\tablecaption{Magnifications at the location of the transient ($\mu$) and 0.06\arcsec\ from the critical curve ($\mu(\theta_c=0.06\arcsec)$).\tablenotemark{a} \label{tab:mag}}
\tablecolumns{4}
\tablewidth{0pt}
\tablehead{
\colhead{Model} &
\colhead{$\mu$} &
\colhead{$\mu(\theta_c=0.06\arcsec)$} &
\colhead{References}
}
\startdata
Brada{\v c} (v3) & 11 & 197 & \cite{hoaghuangtreu16,bradactreuapplegate09,bradacschneiderlombardi05} \\
Caminha (v4) & 56 & 205 & \cite{caminhagrillorosati17} \\
CATS (v4.1) & 15 & 201 & \cite{jauzacclementlimousin14,richardjauzaclimousin14,jauzacjullokneib12} \\
Diego (v4.1) & 25 & 250 & \cite{diegoprotopapassandvik05,diegosandvikprotopapas05,diegotegmarkprotopapas07,diegobroadhurstbenitez15} \\
GLAFIC (v4) & 45 & 180\tablenotemark{b} & \cite{kawamataishigakishimasaku18,kawamataoguriishigaki16,oguri10} \\
Keeton (v4) & 369 & 304 & \cite{mccullykeetonwong14,ammonswongzabludoff14,keeton10} \\
Sharon (v4 Cor.) & 40 & 228 & \cite{johnsonsharonbayliss14,jullokneiblimousin07} \\
Williams/GRALE (v4) & 40\tablenotemark{b} & 250\tablenotemark{b} & 
\cite{sebesta16,liesenborgsderijckedejonghe06} \\
Zitrin-ltm-gauss (v3) & 100 & 331 & \cite{zitrinmeneghettiumetsu13,zitrinbroadhurstumetsu09} (see also \citealt{mertencoedupke11,mertencacciatomeneghetti09}) \\
Zitrin-nfw (v3) & 348 & 208 & \cite{zitrinmeneghettiumetsu13,zitrinbroadhurstumetsu09} (see also \citealt{mertencoedupke11,mertencacciatomeneghetti09}) \\
\enddata
\tablenotetext{a}{Magnifications predicted by MACS J0416.1-2403 lensing models. Those at the transient's coordinates show high dispersion given the uncertainty in the location of the galaxy cluster's critical curve.}
\tablenotetext{b}{Obtained using updated high-resolution maps instead of the published HFF models.}

\end{deluxetable*}


\section{Conclusions}
\label{sec:summary}
In archival {\it HST} imaging taken in 2014 September, we have identified a microlensing event dubbed Warhol in a strongly lensed galaxy at $z=0.94$ very close to the location of the critical curve of the foreground MACS0416 galaxy cluster at $z=0.397$. 
The transient's SED is consistent with the presence of a strong Balmer break, expected for blue supergiant stars, which are also the most common very luminous stars at rest-frame optical wavelengths. 

The lower temperatures and densities of H-rich stellar eruptions, by contrast, generally lack a strong Balmer jump. Further evidence for a microlensing event is the absence of a second detected transient event near the critical curve, as shown in Fig.~\ref{fig:warhol_ims}.  Time delays should be on the order of days at small separations from the critical curve, yet no opposing image is detected. The probability that Warhol could consist of two unresolved images of an outburst is very small, given the comparatively small area in the source plane where any such eruption must occur \citep{kellydiegorodney18}. Warhol's spatial coincidence with the underlying source in the strongly lensed background implies it is very unlikely to be the explosion or outburst of a star in the intracluster medium.

Furthermore, long-term variation in the light curve measured at Warhol's position is consistent with slow fluctuations expected from microlensing by objects in the MACS0416 intracluster medium. 

The frequency of bright microlensing events including Icarus \citep{kellydiegorodney18}, likely the Spock events \citep{rodneybalestrabradac18}, and Warhol provide a new probe of the mass density of objects in the intracluster medium \citep{diegokaiserbroadhurst18,kellydiegorodney18,venumadhavdaimiraldaescude17,oguridiegokaiser18}, as well as the qualitative properties and luminosity functions of massive stars at high redshift \citep{kellydiegorodney18}. 
\citet{diego18} have found that $\sim$50,000 luminous stars at redshifts between $z=1.5$ and $z=2.5$ should experience an average magnification exceeding 100 from lensing halos of all masses. Of these, approximately 8000 stars should have a mean magnification greater than 250 and should exhibit relatively frequent microlensing peaks. \citet{windhorsttimmeswyithe18} have also recently shown that high magnification during caustic-crossing events close to cluster critical curves should provide an opportunity to observe directly Population III stars at high redshifs using the {\it James Webb Space Telescope}.

\startlongtable
\begin{deluxetable}{cccc}
\tablecaption{ Photometry measured from \it{HST} imaging \label{tab:photometry}}
\tablecolumns{4}
\tablewidth{0pt}
\tablehead{
\colhead{Date} &
\colhead{Bandpass} &
\colhead{Flux} &
\colhead{$\sigma$} \\
\colhead{(MJD)} &
\colhead{} &
\colhead{($\mu$Jy)} &
\colhead{($\mu$Jy)}
}
\startdata
56159.53 & ACS F435W & -0.0068 & 0.0129 \\
56184.75 & ACS F435W & -0.0054 & 0.0336 \\
56663.91 & ACS F435W & -0.0054 & 0.0080 \\
56665.62 & ACS F435W & -0.0162 & 0.0109 \\
56668.55 & ACS F435W & 0.0055 & 0.0082 \\
56670.42 & ACS F435W & -0.0049 & 0.0049 \\
56671.94 & ACS F435W & 0.0025 & 0.0083 \\
56672.47 & ACS F435W & 0.0184 & 0.0118 \\
56672.73 & ACS F435W & -0.0123 & 0.0163 \\
56679.25 & ACS F435W & 0.0079 & 0.0099 \\
56686.41 & ACS F435W & 0.0227 & 0.0057 \\
56696.11 & ACS F435W & -0.0008 & 0.0077 \\
56663.43 & ACS F606W & 0.0105 & 0.0083 \\
56665.36 & ACS F606W & 0.0073 & 0.0075 \\
56671.47 & ACS F606W & 0.0040 & 0.0074 \\
56678.25 & ACS F606W & -0.0030 & 0.0071 \\
56682.10 & ACS F606W & -0.0086 & 0.0059 \\
56688.21 & ACS F606W & -0.0098 & 0.0052 \\
56916.89 & ACS F606W & 0.0600 & 0.0109 \\
56184.74 & ACS F814W & 0.0282 & 0.0326 \\
56662.65 & ACS F814W & 0.0062 & 0.0087 \\
56663.56 & ACS F814W & 0.0025 & 0.0122 \\
56664.43 & ACS F814W & -0.0052 & 0.0118 \\
56664.56 & ACS F814W & 0.0199 & 0.0117 \\
56665.49 & ACS F814W & 0.0039 & 0.0152 \\
56666.35 & ACS F814W & -0.0181 & 0.0130 \\
56666.49 & ACS F814W & 0.0325 & 0.0119 \\
56669.28 & ACS F814W & 0.0013 & 0.0128 \\
56670.61 & ACS F814W & -0.0030 & 0.0122 \\
56671.16 & ACS F814W & 0.0078 & 0.0112 \\
56671.60 & ACS F814W & -0.0167 & 0.0113 \\
56672.07 & ACS F814W & -0.0040 & 0.0115 \\
56672.28 & ACS F814W & 0.0166 & 0.0110 \\
56672.60 & ACS F814W & -0.0219 & 0.0103 \\
56672.87 & ACS F814W & -0.0026 & 0.0117 \\
56676.58 & ACS F814W & 0.0050 & 0.0074 \\
56678.91 & ACS F814W & 0.0079 & 0.0124 \\
56680.44 & ACS F814W & -0.0247 & 0.0139 \\
56681.50 & ACS F814W & -0.0053 & 0.0143 \\
56686.15 & ACS F814W & 0.0080 & 0.0124 \\
56686.55 & ACS F814W & 0.0061 & 0.0102 \\
56691.20 & ACS F814W & -0.0146 & 0.0111 \\
56697.44 & ACS F814W & 0.0020 & 0.0120 \\
56697.57 & ACS F814W & 0.0030 & 0.0094 \\
56916.96 & ACS F814W & 0.1072 & 0.0197 \\
56144.84 & WFC3 F105W & 0.0515 & 0.0111 \\
56184.88 & WFC3 F105W & 0.0396 & 0.0279 \\
56689.40 & WFC3 F105W & 0.0072 & 0.0117 \\
56869.77 & WFC3 F105W & 0.0003 & 0.0079 \\
56870.76 & WFC3 F105W & -0.0113 & 0.0111 \\
56877.46 & WFC3 F105W & 0.0009 & 0.0084 \\
56877.73 & WFC3 F105W & -0.0012 & 0.0068 \\
56879.46 & WFC3 F105W & 0.0058 & 0.0046 \\
56880.38 & WFC3 F105W & -0.0065 & 0.0052 \\
56880.65 & WFC3 F105W & -0.0033 & 0.0077 \\
56881.71 & WFC3 F105W & -0.0110 & 0.0089 \\
56889.81 & WFC3 F105W & 0.0102 & 0.0067 \\
56898.77 & WFC3 F105W & 0.0028 & 0.0068 \\
56899.04 & WFC3 F105W & 0.0016 & 0.0061 \\
56900.10 & WFC3 F105W & 0.0090 & 0.0061 \\
56984.57 & WFC3 F105W & 0.0385 & 0.0327 \\
56991.60 & WFC3 F105W & 0.0033 & 0.0347 \\
57035.58 & WFC3 F105W & -0.0284 & 0.0316 \\
57040.55 & WFC3 F105W & 0.0131 & 0.0443 \\
56159.60 & WFC3 F125W & 0.0065 & 0.0185 \\
56197.79 & WFC3 F125W & 0.0149 & 0.0397 \\
56689.34 & WFC3 F125W & -0.0026 & 0.0157 \\
56871.04 & WFC3 F125W & -0.0049 & 0.0075 \\
56876.93 & WFC3 F125W & 0.0033 & 0.0048 \\
56897.84 & WFC3 F125W & -0.0003 & 0.0062 \\
56899.97 & WFC3 F125W & -0.0043 & 0.0062 \\
56900.64 & WFC3 F125W & 0.0059 & 0.0069 \\
56901.83 & WFC3 F125W & 0.0000 & 0.0079 \\
56915.76 & WFC3 F125W & 0.1121 & 0.0131 \\
56922.33 & WFC3 F125W & 0.0291 & 0.0247 \\
56928.05 & WFC3 F125W & -0.0121 & 0.0242 \\
56159.62 & WFC3 F140W & 0.0238 & 0.0280 \\
56184.87 & WFC3 F140W & 0.0489 & 0.0201 \\
56874.94 & WFC3 F140W & 0.0006 & 0.0048 \\
56875.87 & WFC3 F140W & 0.0016 & 0.0058 \\
56888.95 & WFC3 F140W & 0.0125 & 0.0060 \\
56890.67 & WFC3 F140W & -0.0133 & 0.0057 \\
56899.84 & WFC3 F140W & 0.0008 & 0.0077 \\
56984.63 & WFC3 F140W & 0.0417 & 0.0240 \\
56991.47 & WFC3 F140W & 0.0169 & 0.0372 \\
57035.45 & WFC3 F140W & -0.0287 & 0.0449 \\
57040.62 & WFC3 F140W & -0.0187 & 0.0293 \\
56132.22 & WFC3 F160W & 0.0276 & 0.0256 \\
56144.86 & WFC3 F160W & 0.0471 & 0.0376 \\
56170.77 & WFC3 F160W & 0.0496 & 0.0191 \\
56197.77 & WFC3 F160W & 0.0342 & 0.0241 \\
56689.33 & WFC3 F160W & 0.0378 & 0.0212 \\
56869.78 & WFC3 F160W & -0.0124 & 0.0099 \\
56870.78 & WFC3 F160W & -0.0122 & 0.0081 \\
56877.48 & WFC3 F160W & -0.0225 & 0.0141 \\
56877.75 & WFC3 F160W & 0.0001 & 0.0091 \\
56879.47 & WFC3 F160W & -0.0030 & 0.0075 \\
56880.40 & WFC3 F160W & 0.0034 & 0.0090 \\
56880.67 & WFC3 F160W & 0.0125 & 0.0075 \\
56881.73 & WFC3 F160W & -0.0122 & 0.0126 \\
56889.83 & WFC3 F160W & 0.0047 & 0.0066 \\
56898.79 & WFC3 F160W & 0.0101 & 0.0125 \\
56899.06 & WFC3 F160W & 0.0168 & 0.0088 \\
56900.12 & WFC3 F160W & 0.0151 & 0.0086 \\
56915.70 & WFC3 F160W & 0.1311 & 0.0163 \\
56922.39 & WFC3 F160W & 0.0808 & 0.0323 \\
56928.12 & WFC3 F160W & 0.0153 & 0.0268
\enddata
\end{deluxetable}

\acknowledgements
This work was supported in part by World Premier International Research Center Initiative (WPI Initiative), MEXT, Japan, and JSPS KAKENHI grants JP15H05892 and JP18K03693.
R.J.F.\ is supported in part by NSF grant AST-1518052, the Gordon \& Betty Moore Foundation, the Heising-Simons
Foundation, and by a fellowship from the David and Lucile Packard Foundation.  A.V.F. and T.T. acknowledge generous financial assistance from NASA/{\it HST} grants GO-14922 and GO-14872 from the Space Telescope Science Institute (STScI), which is  operated by the Association of Universities for Research in Astronomy, Inc., under NASA contract NAS 5-26555. Additional support for A.V.F. was provided by the Christopher R.\ Redlich Fund, the TABASGO Foundation,
and the Miller Institute for Basic Research in Science (U.C. Berkeley). J.M.D. acknowledges the support of projects AYA2015-64508-P (MINECO/FEDER, UE), funded by the Ministerio de Economia y Competitividad. 
J.H. was supported by a VILLUM FONDEN Investigator grant (project number 16599).
This work utilizes gravitational lensing models produced by PIs Bradač, Natarajan, \& Kneib (CATS); Merten \& Zitrin; Sharon, Williams, Keeton, Bernstein, and Diego; and the GLAFIC group. This lens modeling was partially funded by the {\it HST} Frontier Fields program conducted by STScI. The lens models were obtained from the Mikulski Archive for Space Telescopes (MAST).

\bibliography{apj_paper}

\begin{thebibliography}{}
\expandafter\ifx\csname natexlab\endcsname\relax\def\natexlab#1{#1}\fi

\bibitem[{{Ammons} {et~al.}(2014){Ammons}, {Wong}, {Zabludoff}, \&
  {Keeton}}]{ammonswongzabludoff14}
{Ammons}, S.~M., {Wong}, K.~C., {Zabludoff}, A.~I., \& {Keeton}, C.~R. 2014,
  \apj, 781, 2

\bibitem[{{Balestra} {et~al.}(2016){Balestra}, {Mercurio}, {Sartoris},
  {Girardi}, {Grillo}, {Nonino}, {Rosati}, {Biviano}, {Ettori}, {Forman},
  {Jones}, {Koekemoer}, {Medezinski}, {Merten}, {Ogrean}, {Tozzi}, {Umetsu},
  {Vanzella}, {van Weeren}, {Zitrin}, {Annunziatella}, {Caminha}, {Broadhurst},
  {Coe}, {Donahue}, {Fritz}, {Frye}, {Kelson}, {Lombardi}, {Maier},
  {Meneghetti}, {Monna}, {Postman}, {Scodeggio}, {Seitz}, \&
  {Ziegler}}]{balestramercuriosartoris16}
{Balestra}, I., {Mercurio}, A., {Sartoris}, B., {et~al.} 2016, \apjs, 224, 33

\bibitem[{{Brada{\v c}} {et~al.}(2005){Brada{\v c}}, {Schneider}, {Lombardi},
  \& {Erben}}]{bradacschneiderlombardi05}
{Brada{\v c}}, M., {Schneider}, P., {Lombardi}, M., \& {Erben}, T. 2005, \aap,
  437, 39

\bibitem[{{Brada{\v c}} {et~al.}(2009){Brada{\v c}}, {Treu}, {Applegate},
  {Gonzalez}, {Clowe}, {Forman}, {Jones}, {Marshall}, {Schneider}, \&
  {Zaritsky}}]{bradactreuapplegate09}
{Brada{\v c}}, M., {Treu}, T., {Applegate}, D., {et~al.} 2009, \apj, 706, 1201

\bibitem[{{Caminha} {et~al.}(2017){Caminha}, {Grillo}, {Rosati}, {Balestra},
  {Mercurio}, {Vanzella}, {Biviano}, {Caputi}, {Delgado-Correal}, {Karman},
  {Lombardi}, {Meneghetti}, {Sartoris}, \& {Tozzi}}]{caminhagrillorosati17}
{Caminha}, G.~B., {Grillo}, C., {Rosati}, P., {et~al.} 2017, \aap, 600, A90

\bibitem[{{Castelli} \& {Kurucz}(2004)}]{castellikurucz04}
{Castelli}, F., \& {Kurucz}, R.~L. 2004, ArXiv Astrophysics e-prints,
  astro-ph/0405087

\bibitem[{{Dachs}(1970)}]{dachs70}
{Dachs}, J. 1970, \aap, 9, 95

\bibitem[{{Diego}(2018)}]{diego18}
{Diego}, J.~M. 2018, ArXiv e-prints, arXiv:1806.04668

\bibitem[{{Diego} {et~al.}(2005{\natexlab{a}}){Diego}, {Protopapas}, {Sandvik},
  \& {Tegmark}}]{diegoprotopapassandvik05}
{Diego}, J.~M., {Protopapas}, P., {Sandvik}, H.~B., \& {Tegmark}, M.
  2005{\natexlab{a}}, \mnras, 360, 477

\bibitem[{{Diego} {et~al.}(2005{\natexlab{b}}){Diego}, {Sandvik}, {Protopapas},
  {Tegmark}, {Ben{\'{\i}}tez}, \& {Broadhurst}}]{diegosandvikprotopapas05}
{Diego}, J.~M., {Sandvik}, H.~B., {Protopapas}, P., {et~al.}
  2005{\natexlab{b}}, \mnras, 362, 1247

\bibitem[{{Diego} {et~al.}(2007){Diego}, {Tegmark}, {Protopapas}, \&
  {Sandvik}}]{diegotegmarkprotopapas07}
{Diego}, J.~M., {Tegmark}, M., {Protopapas}, P., \& {Sandvik}, H.~B. 2007,
  \mnras, 375, 958

\bibitem[{{Diego} {et~al.}(2015){Diego}, {Broadhurst}, {Benitez}, {Umetsu},
  {Coe}, {Sendra}, {Sereno}, {Izzo}, \& {Covone}}]{diegobroadhurstbenitez15}
{Diego}, J.~M., {Broadhurst}, T., {Benitez}, N., {et~al.} 2015, \mnras, 446,
  683

\bibitem[{{Diego} {et~al.}(2018){Diego}, {Kaiser}, {Broadhurst}, {Kelly},
  {Rodney}, {Morishita}, {Oguri}, {Ross}, {Zitrin}, {Jauzac}, {Richard},
  {Williams}, {Vega-Ferrero}, {Frye}, \&
  {Filippenko}}]{diegokaiserbroadhurst18}
{Diego}, J.~M., {Kaiser}, N., {Broadhurst}, T., {et~al.} 2018, \apj, 857, 25

\bibitem[{{Doran} {et~al.}(2013){Doran}, {Crowther}, {de Koter}, {Evans},
  {McEvoy}, {Walborn}, {Bastian}, {Bestenlehner}, {Gr{\"a}fener}, {Herrero},
  {K{\"o}hler}, {Ma{\'\i}z Apell{\'a}niz}, {Najarro}, {Puls}, {Sana},
  {Schneider}, {Taylor}, {van Loon}, \& {Vink}}]{dorancrowtherdekoter13}
{Doran}, E.~I., {Crowther}, P.~A., {de Koter}, A., {et~al.} 2013, \aap, 558,
  A134

\bibitem[{{Ebeling} {et~al.}(2001){Ebeling}, {Edge}, \&
  {Henry}}]{ebelingedgehenry01}
{Ebeling}, H., {Edge}, A.~C., \& {Henry}, J.~P. 2001, \apj, 553, 668

\bibitem[{{Fruchter} {et~al.}(2010){Fruchter}, {Hack}, {Dencheva},
  {Droettboom}, \& {Greenfield}}]{fruchter10}
{Fruchter}, A.~S., {Hack}, W., {Dencheva}, M., {Droettboom}, M., \&
  {Greenfield}, P. 2010, in 2010 Space Telescope Science Institute Calibration
  Workshop, p. 382-387, 382--387

\bibitem[{{Hoag} {et~al.}(2016){Hoag}, {Huang}, {Treu}, {Brada{\v c}},
  {Schmidt}, {Wang}, {Brammer}, {Broussard}, {Amorin}, {Castellano}, {Fontana},
  {Merlin}, {Schrabback}, {Trenti}, \& {Vulcani}}]{hoaghuangtreu16}
{Hoag}, A., {Huang}, K.-H., {Treu}, T., {et~al.} 2016, \apj, 831, 182

\bibitem[{{Jauzac} {et~al.}(2012){Jauzac}, {Jullo}, {Kneib}, {Ebeling},
  {Leauthaud}, {Ma}, {Limousin}, {Massey}, \& {Richard}}]{jauzacjullokneib12}
{Jauzac}, M., {Jullo}, E., {Kneib}, J.-P., {et~al.} 2012, \mnras, 426, 3369

\bibitem[{{Jauzac} {et~al.}(2014){Jauzac}, {Cl{\'e}ment}, {Limousin},
  {Richard}, {Jullo}, {Ebeling}, {Atek}, {Kneib}, {Knowles}, {Natarajan},
  {Eckert}, {Egami}, {Massey}, \& {Rexroth}}]{jauzacclementlimousin14}
{Jauzac}, M., {Cl{\'e}ment}, B., {Limousin}, M., {et~al.} 2014, \mnras, 443,
  1549

\bibitem[{{Johnson} {et~al.}(2014){Johnson}, {Sharon}, {Bayliss}, {Gladders},
  {Coe}, \& {Ebeling}}]{johnsonsharonbayliss14}
{Johnson}, T.~L., {Sharon}, K., {Bayliss}, M.~B., {et~al.} 2014, \apj, 797, 48

\bibitem[{{Jones} {et~al.}(2015){Jones}, {Scolnic}, \&
  {Rodney}}]{jonesscolnicrodney15}
{Jones}, D.~O., {Scolnic}, D.~M., \& {Rodney}, S.~A. 2015, {PythonPhot: Simple
  DAOPHOT-type photometry in Python}, {\it Astrophys. Source Code Lib.},
  ascl:1501.010

\bibitem[{{Jullo} {et~al.}(2007){Jullo}, {Kneib}, {Limousin},
  {El{\'{\i}}asd{\'o}ttir}, {Marshall}, \& {Verdugo}}]{jullokneiblimousin07}
{Jullo}, E., {Kneib}, J.-P., {Limousin}, M., {et~al.} 2007, New J. Phys., 9,
  447

\bibitem[{{Kawamata} {et~al.}(2018){Kawamata}, {Ishigaki}, {Shimasaku},
  {Oguri}, {Ouchi}, \& {Tanigawa}}]{kawamataishigakishimasaku18}
{Kawamata}, R., {Ishigaki}, M., {Shimasaku}, K., {et~al.} 2018, \apj, 855, 4

\bibitem[{{Kawamata} {et~al.}(2016){Kawamata}, {Oguri}, {Ishigaki},
  {Shimasaku}, \& {Ouchi}}]{kawamataoguriishigaki16}
{Kawamata}, R., {Oguri}, M., {Ishigaki}, M., {Shimasaku}, K., \& {Ouchi}, M.
  2016, \apj, 819, 114

\bibitem[{{Keeton}(2010)}]{keeton10}
{Keeton}, C.~R. 2010, Gen. Rel. Grav., 42, 2151

\bibitem[{{Kelly} {et~al.}(2018){Kelly}, {Diego}, {Rodney}, {Kaiser},
  {Broadhurst}, {Zitrin}, {Treu}, {P{\'e}rez-Gonz{\'a}lez}, {Morishita},
  {Jauzac}, {Selsing}, {Oguri}, {Pueyo}, {Ross}, {Filippenko}, {Smith},
  {Hjorth}, {Cenko}, {Wang}, {Howell}, {Richard}, {Frye}, {Jha}, {Foley},
  {Norman}, {Bradac}, {Zheng}, {Brammer}, {Benito}, {Cava}, {Christensen}, {de
  Mink}, {Graur}, {Grillo}, {Kawamata}, {Kneib}, {Matheson}, {McCully},
  {Nonino}, {P{\'e}rez-Fournon}, {Riess}, {Rosati}, {Schmidt}, {Sharon}, \&
  {Weiner}}]{kellydiegorodney18}
{Kelly}, P.~L., {Diego}, J.~M., {Rodney}, S., {et~al.} 2018, Nature Astronomy,
  2, 334

\bibitem[{{Kim} {et~al.}(1996){Kim}, {Goobar}, \&
  {Perlmutter}}]{kimgoobarperlmutter96}
{Kim}, A., {Goobar}, A., \& {Perlmutter}, S. 1996, \pasp, 108, 190

\bibitem[{{Liesenborgs} {et~al.}(2006){Liesenborgs}, {De Rijcke}, \&
  {Dejonghe}}]{liesenborgsderijckedejonghe06}
{Liesenborgs}, J., {De Rijcke}, S., \& {Dejonghe}, H. 2006, \mnras, 367, 1209

\bibitem[{{Lotz} {et~al.}(2017){Lotz}, {Koekemoer}, {Coe}, {Grogin}, {Capak},
  {Mack}, {Anderson}, {Avila}, {Barker}, {Borncamp}, {Brammer}, {Durbin},
  {Gunning}, {Hilbert}, {Jenkner}, {Khandrika}, {Levay}, {Lucas}, {MacKenty},
  {Ogaz}, {Porterfield}, {Reid}, {Robberto}, {Royle}, {Smith},
  {Storrie-Lombardi}, {Sunnquist}, {Surace}, {Taylor}, {Williams}, {Bullock},
  {Dickinson}, {Finkelstein}, {Natarajan}, {Richard}, {Robertson}, {Tumlinson},
  {Zitrin}, {Flanagan}, {Sembach}, {Soifer}, \&
  {Mountain}}]{lotzkoekemoercoe17}
{Lotz}, J.~M., {Koekemoer}, A., {Coe}, D., {et~al.} 2017, \apj, 837, 97

\bibitem[{{McCully} {et~al.}(2014){McCully}, {Keeton}, {Wong}, \&
  {Zabludoff}}]{mccullykeetonwong14}
{McCully}, C., {Keeton}, C.~R., {Wong}, K.~C., \& {Zabludoff}, A.~I. 2014,
  \mnras, 443, 3631

\bibitem[{{Merten} {et~al.}(2009){Merten}, {Cacciato}, {Meneghetti}, {Mignone},
  \& {Bartelmann}}]{mertencacciatomeneghetti09}
{Merten}, J., {Cacciato}, M., {Meneghetti}, M., {Mignone}, C., \& {Bartelmann},
  M. 2009, \aap, 500, 681

\bibitem[{{Merten} {et~al.}(2011){Merten}, {Coe}, {Dupke}, {Massey}, {Zitrin},
  {Cypriano}, {Okabe}, {Frye}, {Braglia}, {Jim{\'e}nez-Teja}, {Ben{\'{\i}}tez},
  {Broadhurst}, {Rhodes}, {Meneghetti}, {Moustakas}, {Sodr{\'e}}, {Krick}, \&
  {Bregman}}]{mertencoedupke11}
{Merten}, J., {Coe}, D., {Dupke}, R., {et~al.} 2011, \mnras, 417, 333

\bibitem[{{Miralda-Escude}(1991)}]{miraldaescude91}
{Miralda-Escude}, J. 1991, \apj, 379, 94

\bibitem[{{Oguri}(2010)}]{oguri10}
{Oguri}, M. 2010, \pasj, 62, 1017

\bibitem[{{Oguri} {et~al.}(2018){Oguri}, {Diego}, {Kaiser}, {Kelly}, \&
  {Broadhurst}}]{oguridiegokaiser18}
{Oguri}, M., {Diego}, J.~M., {Kaiser}, N., {Kelly}, P.~L., \& {Broadhurst}, T.
  2018, \prd, 97, 023518

\bibitem[{{Oke} \& {Gunn}(1983)}]{okegunn83}
{Oke}, J.~B., \& {Gunn}, J.~E. 1983, \apj, 266, 713

\bibitem[{{Patr{\'{\i}}cio} {et~al.}(2018){Patr{\'{\i}}cio}, {Richard},
  {Carton}, {Contini}, {Epinat}, {Brinchmann}, {Schmidt}, {Krajnovi{\'c}},
  {Bouch{\'e}}, {Weilbacher}, {Pell{\'o}}, {Caruana}, {Maseda}, {Finley},
  {Bauer}, {Martinez}, {Mahler}, {Lagattuta}, {Cl{\'e}ment}, {Soucail}, \&
  {Wisotzki}}]{patriciorichardcarton18}
{Patr{\'{\i}}cio}, V., {Richard}, J., {Carton}, D., {et~al.} 2018, \mnras, 477,
  18

\bibitem[{{Pickles}(1998)}]{pickles98}
{Pickles}, A.~J. 1998, \pasp, 110, 863

\bibitem[{{Postman} {et~al.}(2012){Postman}, {Coe}, {Ben{\'{\i}}tez},
  {Bradley}, {Broadhurst}, {Donahue}, {Ford}, {Graur}, {Graves}, {Jouvel},
  {Koekemoer}, {Lemze}, {Medezinski}, {Molino}, {Moustakas}, {Ogaz}, {Riess},
  {Rodney}, {Rosati}, {Umetsu}, {Zheng}, {Zitrin}, {Bartelmann}, {Bouwens},
  {Czakon}, {Golwala}, {Host}, {Infante}, {Jha}, {Jimenez-Teja}, {Kelson},
  {Lahav}, {Lazkoz}, {Maoz}, {McCully}, {Melchior}, {Meneghetti}, {Merten},
  {Moustakas}, {Nonino}, {Patel}, {Reg{\"o}s}, {Sayers}, {Seitz}, \& {Van der
  Wel}}]{postmancoebenitez12}
{Postman}, M., {Coe}, D., {Ben{\'{\i}}tez}, N., {et~al.} 2012, \apjs, 199, 25

\bibitem[{{Richard} {et~al.}(2014){Richard}, {Jauzac}, {Limousin}, {Jullo},
  {Cl{\'e}ment}, {Ebeling}, {Kneib}, {Atek}, {Natarajan}, {Egami}, {Livermore},
  \& {Bower}}]{richardjauzaclimousin14}
{Richard}, J., {Jauzac}, M., {Limousin}, M., {et~al.} 2014, \mnras, 444, 268

\bibitem[{{Rodney} {et~al.}(2018){Rodney}, {Balestra}, {Bradac}, {Brammer},
  {Broadhurst}, {Caminha}, {Chiriv{\i}}, {Diego}, {Filippenko}, {Foley},
  {Graur}, {Grillo}, {Hemmati}, {Hjorth}, {Hoag}, {Jauzac}, {Jha}, {Kawamata},
  {Kelly}, {McCully}, {Mobasher}, {Molino}, {Oguri}, {Richard}, {Riess},
  {Rosati}, {Schmidt}, {Selsing}, {Sharon}, {Strolger}, {Suyu}, {Treu},
  {Weiner}, {Williams}, \& {Zitrin}}]{rodneybalestrabradac18}
{Rodney}, S.~A., {Balestra}, I., {Bradac}, M., {et~al.} 2018, Nature Astronomy,
  2, 324

\bibitem[{{Schlafly} \& {Finkbeiner}(2011)}]{schlaflyfinkbeinerSFD11}
{Schlafly}, E.~F., \& {Finkbeiner}, D.~P. 2011, \apj, 737, 103

\bibitem[{{Schmidt} {et~al.}(2014){Schmidt}, {Treu}, {Brammer}, {Brada{\v c}},
  {Wang}, {Dijkstra}, {Dressler}, {Fontana}, {Gavazzi}, {Henry}, {Hoag},
  {Jones}, {Kelly}, {Malkan}, {Mason}, {Pentericci}, {Poggianti}, {Stiavelli},
  {Trenti}, {von der Linden}, \& {Vulcani}}]{schmidttreubrammer14}
{Schmidt}, K.~B., {Treu}, T., {Brammer}, G.~B., {et~al.} 2014, \apjl, 782, L36

\bibitem[{{Sebesta} {et~al.}(2016){Sebesta}, {Williams}, {Mohammed}, {Saha}, \&
  {Liesenborgs}}]{sebesta16}
{Sebesta}, K., {Williams}, L.~L.~R., {Mohammed}, I., {Saha}, P., \&
  {Liesenborgs}, J. 2016, \mnras, 461, 2126

\bibitem[{{Stetson}(1987)}]{stetson87}
{Stetson}, P.~B. 1987, \pasp, 99, 191

\bibitem[{{Treu} {et~al.}(2015){Treu}, {Schmidt}, {Brammer}, {Vulcani}, {Wang},
  {Brada{\v c}}, {Dijkstra}, {Dressler}, {Fontana}, {Gavazzi}, {Henry}, {Hoag},
  {Huang}, {Jones}, {Kelly}, {Malkan}, {Mason}, {Pentericci}, {Poggianti},
  {Stiavelli}, {Trenti}, \& {von der Linden}}]{treuschmidtbrammer15}
{Treu}, T., {Schmidt}, K.~B., {Brammer}, G.~B., {et~al.} 2015, \apj, 812, 114

\bibitem[{{Venumadhav} {et~al.}(2017){Venumadhav}, {Dai}, \&
  {Miralda-Escud{\'e}}}]{venumadhavdaimiraldaescude17}
{Venumadhav}, T., {Dai}, L., \& {Miralda-Escud{\'e}}, J. 2017, \apj, 850, 49

\bibitem[{{Windhorst} {et~al.}(2018){Windhorst}, {Timmes}, {Wyithe},
  {Alpaslan}, {Andrews}, {Coe}, {Diego}, {Dijkstra}, {Driver}, {Kelly}, \&
  {Kim}}]{windhorsttimmeswyithe18}
{Windhorst}, R.~A., {Timmes}, F.~X., {Wyithe}, J.~S.~B., {et~al.} 2018, \apjs,
  234, 41

\bibitem[{{Zitrin} {et~al.}(2009){Zitrin}, {Broadhurst}, {Umetsu}, {Coe},
  {Ben{\'{\i}}tez}, {Ascaso}, {Bradley}, {Ford}, {Jee}, {Medezinski},
  {Rephaeli}, \& {Zheng}}]{zitrinbroadhurstumetsu09}
{Zitrin}, A., {Broadhurst}, T., {Umetsu}, K., {et~al.} 2009, \mnras, 396, 1985

\bibitem[{{Zitrin} {et~al.}(2013){Zitrin}, {Meneghetti}, {Umetsu},
  {Broadhurst}, {Bartelmann}, {Bouwens}, {Bradley}, {Carrasco}, {Coe}, {Ford},
  {Kelson}, {Koekemoer}, {Medezinski}, {Moustakas}, {Moustakas}, {Nonino},
  {Postman}, {Rosati}, {Seidel}, {Seitz}, {Sendra}, {Shu}, {Vega}, \&
  {Zheng}}]{zitrinmeneghettiumetsu13}
{Zitrin}, A., {Meneghetti}, M., {Umetsu}, K., {et~al.} 2013, \apjl, 762, L30

\end{thebibliography}

\end{document}